\documentclass[twoside,a4paper,11pt]{sea10}
\usepackage{graphicx}
\usepackage{hyperref}
\usepackage{movie15}
\begin{document}
\pagenumbering{arabic}
\pagestyle{myheadings}
\thispagestyle{empty}
{\flushleft\includegraphics[width=\textwidth,bb=58 650 590 680]{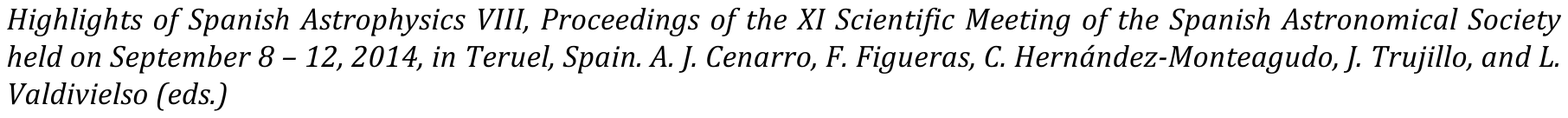}}
\vspace*{0.2cm}
\begin{flushleft}
{\bf {\LARGE
%
%%% TITLE of the paper. 
%%% TITLE of the paper. 
The nature of extragalactic radio-jets from high-resolution radio-interferometric observations.
%
% Do not delete next few lines
}\\
\vspace*{1cm}
%
%%% Include here the LIST OF AUTHORS.
%%% Include here the LIST OF AUTHORS.
%%% Note that the last author has to be preceeded by an AND.
Manel Perucho-Pla$^{1,2}$.
%
% Do not delete next few lines
}\\
\vspace*{0.5cm}
%
%%% AFFILIATIONS LIST.
%%% and the AFFILIATIONS LIST. Note that one affiliation per line.
%%% Add as many affiliations as necessary. 
$^{1}$
Departament d'Astronomia i Astrof\'{\i}sica. Universitat de Val\`encia. Av. Vicent Andr\'es Estell\'es s/n, 46100 Burjassot (Val\`encia).\\
$^{2}$
Observatori Astron\`omic, Parc Cient\'{\i}fic, Universitat de Val\`encia, C/ Catedr\`atic Jos\'e Beltr\'an 2, 46980 Paterna (Val\`encia).%
% Do not delete next few lines
\end{flushleft}
%
% Headings
\markboth{
%%% Type the SHORT version of the paper title.
%%% Type the SHORT version of the paper title.
On the nature of extragalactic radio-jets.
}{ % Do not delete
%
%%%  First Author \& Second Author   OR   First-author et al. 
%%%  First Author \& Second Author   OR   First-author et al. if the author list 
%%% contains three or more authors.
Manel Perucho 
% 
% Do not delete next few lines
}
\thispagestyle{empty}
\vspace*{0.4cm}
\begin{minipage}[l]{0.09\textwidth}
\ 
\end{minipage}
\begin{minipage}[r]{0.9\textwidth}
\vspace{1cm}
\section*{Abstract}{\small
%
% ABSTRACT ABSTRACT ABSTRACT
% ABSTRACT ABSTRACT ABSTRACT
%%% Type the ABSTRACT of your paper
Extragalactic jets are a common feature of radio-loud active galaxies. The nature of the observed jets in relation to the bulk flow is still unclear. In particular it is not clear whether the observations of parsec-scale jets using the very long baseline interferometric technique (VLBI) reveal wave-like structures that develop and propagate along the jet, or trace the jet flow itself. In this contribution I review the evidence collected during the last years showing that the ridge-lines of helical radio-jets do not correspond to observational artifacts. 
This conclusion was reached by studying a number of VLBI observations of the radio jet in the quasar S5~0836+710 at different frequencies and epochs. 
The ridge-line of the emission in the jet coincides at all frequencies within the errors. Moreover, small differences between the ridge-lines as observed at different epochs reveal wave-like motion transversal to the jet propagation axis. I also discuss similar results, albeit with different interpretations, obtained by other authors. The current challenge is to measure the propagation velocities of these waves and to try to characterise them in terms of simple perturbations or Kelvin-Helmholtz instability, which would help understanding the physical conditions of the flow where the waves develop. This problem can only be tackled by high-resolution observations such as those that can be achieved by the space radio-antenna \emph{Radioastron}.
%
% Do not delete next few lines
\normalsize}
\end{minipage}
%
%
%%% BODY of the paper
%%% BODY of the paper
%
\section{Introduction \label{intro}}

 Extragalactic jets are a common feature of radio-loud active galaxies \cite{bhk13}. These galaxies presumably host a supermassive black hole in their nuclei ($M_{\rm BH}=10^8-10^{10}\,M_{\odot}$), and the accretion disks of matter that form around them originate the observed strong optical/UV emission. Around one per cent of active galaxies are radio-loud, with the radio emission being emitted from jets. Current models suggest that the strong magnetic fields in these systems are able to extract energy from the black-hole or the accretion disk \cite{bz77,bp82}. The particles get collimated and accelerated by the magnetic fields, acquiring terminal speeds close to the speed of light \cite{ko09}. Jets propagate through the host galaxy and even outside, bringing large amounts of particles and energy from the central regions of the host galaxies to regions hundreds of kiloparsecs away ($\sim 10^{61-62}\,{\rm erg}$ during $10^8\,{\rm yr}$ for powerful jets). This process completely alters the gas distribution and evolution of the host galaxies (typically massive ellipticals at the centers of clusters), as a direct consequence of the crossing of a supersonic flow. Thus, it is important to understand the nature and properties of jets not only because they are macroscopic laboratories of plasma physics but also to derive accurate information about the processes that determine their interaction with the ambient and quantify the role of relativistic outflows in galaxy and cluster evolution. 
 
  Jets emit electromagnetic radiation throughout the whole electromagnetic spectrum, primarily via synchrotron and inverse Compton radiation. The parsec-scale structure of jets in active galactic nuclei (AGN) is mainly observed in the radio band, using the 
VLBI technique. The nature and properties of the emitting region as related to 
the flow are still scarcely known. Jets can be treated as flows because the Larmor radius of particles is very small 
when compared to the spatial scales of the problem studied \cite{br74}, thus allowing to consider the whole system as a fluid. This interpretation has different implications: Such systems, which are composed of magnetic fields and particles propagating along the jet channel are expected to host the growth of different hydrodynamical and/or magnetic instabilities.  The instabilities are waves triggered by any external or internal perturbation that grow in amplitude with distance, as they are advected with the flow (see reviews in refs.~\cite{bi91,ha06,ha11,pe12}). The jet acts in a similar way to waveguides, hosting the waves that bounce against its ``walls''. In the case of the unstable modes the reflection coefficient of the wave is larger than one \cite{pc85}. The energy gain comes from the kinetic energy of the flow, which is correspondingly decelerated. If instabilities do really grow in jets, we would expect to observe wave-like structures. Different works have related observed structures to instabilities in extragalactic jets, at parsec-scales (knots, bendings, helices; see e.g., \cite{lo98,lz01,ha05,lo06,pl07}). 

  Current issues that are open in jet physics include an aspect as basic as the nature of the observed radio jets in active galaxies. In particular it is not clear whether the VLBI observations of parsec-scale jets reveal wave-like structures that develop and propagate along the jet, or they trace the whole jet flow. Whether the jet flow is energetically dominated by non-radiating thermal particles or it is magnetically dominated up to kiloparsec scales, as expected to be the case at the formation region, is also a matter of debate.

    In reference \cite{pe12a}, the authors used observations of the jet in S5~0836+710 at different frequencies and epochs and showed that the ridge line of this jet behaves as expected if they are interpreted as pressure waves. The jet in BL~Lac also shows wave-like structures propagating along the jet and these are identified by the authors as Alfv\'en waves in a magnetically dominated jet \cite{co14a,co14b}. In this contribution I review results obtained for the quasar S5~0836+710 using VLBI observations and put them in context with other recent works.

  The luminous quasar S5~0836+710 is at a redshift $z=2.16$ and hosts a powerful radio jet extending up to kiloparsec scales \cite{hu92}. At this redshift, $1\,\rm{mas} \simeq 8.4\,\rm{pc}$ (see, e.g., MOJAVE database). VLBI monitoring of the
source showed kink structures \cite{kr90} and yielded estimates of the bulk Lorentz factor
$\gamma_\mathrm{j}=12$ and the viewing angle $\alpha_\mathrm{j}=3^\circ$ of the
flow at milliarcsecond scales \cite{ot98}. The jet was observed
at 1.6 and 5~GHz with VSOP (VLBI Space Observatory Program, a Japanese-led 
space VLBI mission), and oscillations of the ridge-line were also observed \cite{lo98,lo06}.
These oscillations were interpreted as produced by Kelvin-Helmholtz (KH) instability modes \cite{lo06}. The source has also been monitored by the MOJAVE program \cite{li09}. Finally, it has been suggested that the lack of a collimated jet structure and hot-spot at the arc-second scales is due to jet disruption, as indicated by the growth in the amplitude of the helical structure \cite{pe12b}.

 \begin{figure}
\center
\includegraphics[width=0.45\textwidth,angle=0,clip=true]{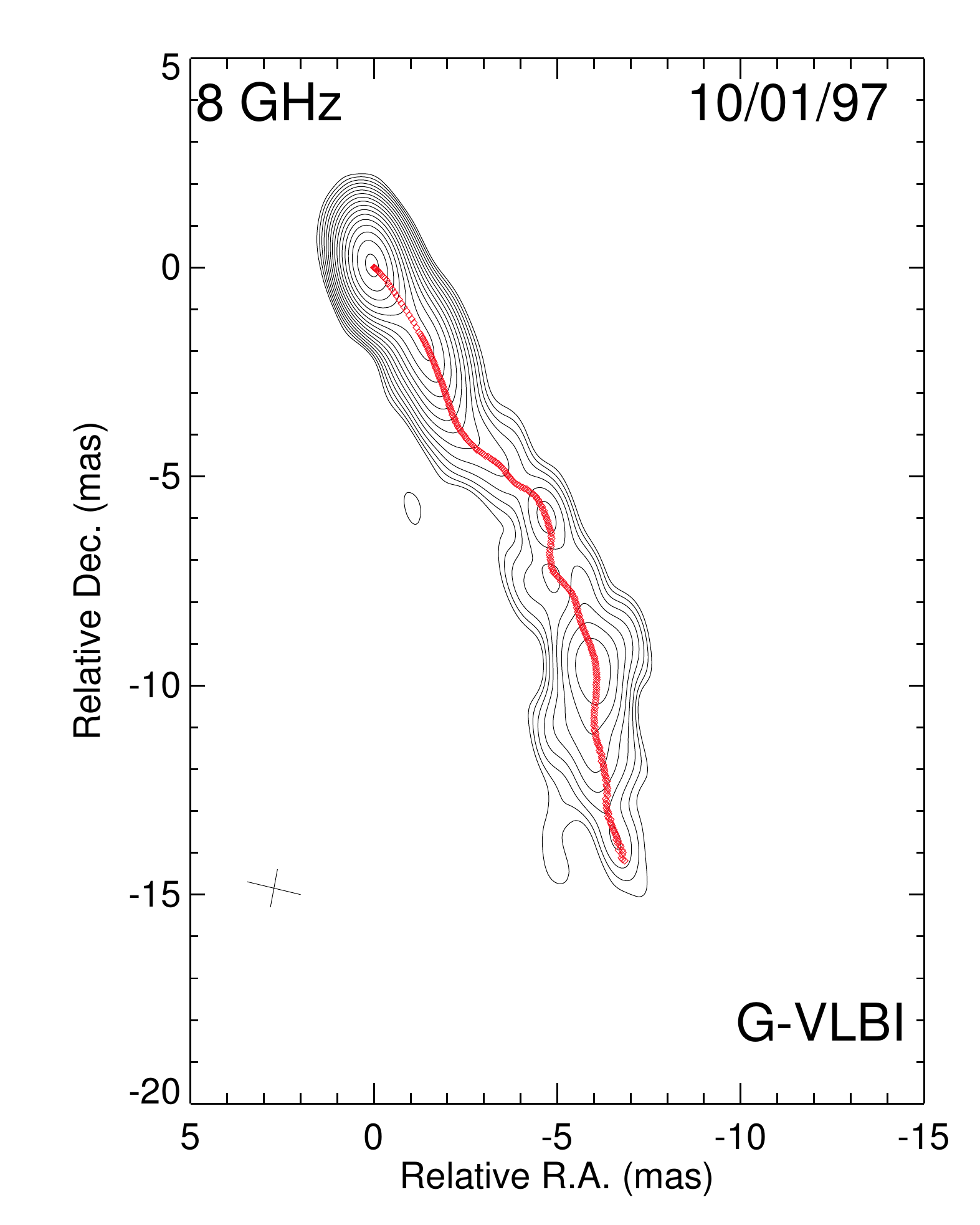} 
\includegraphics[width=0.45\textwidth,angle=0,clip=true]{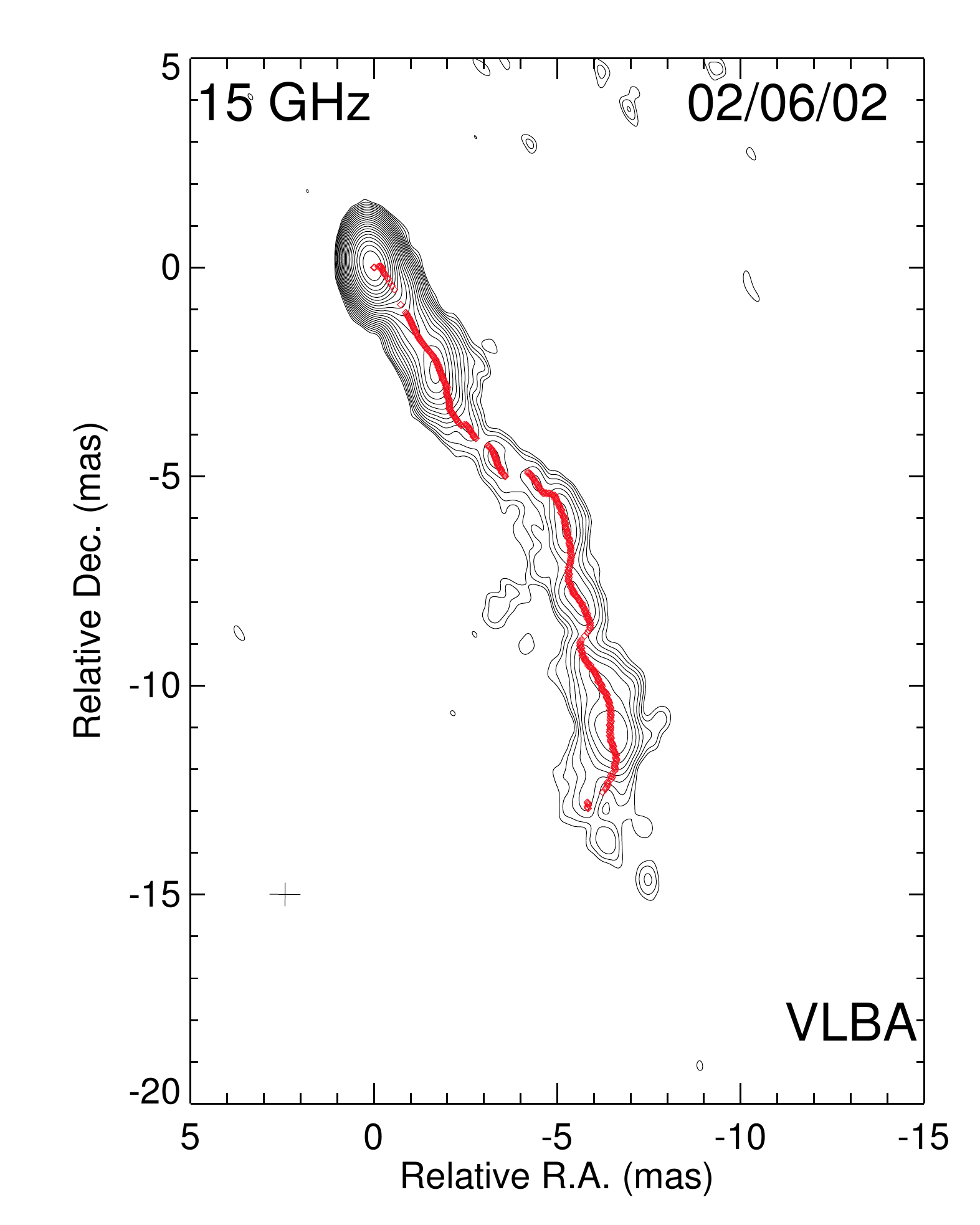} 
\caption{\label{fig1} The jet in S5~0836+710 as obtained with Global VLBI and VLBA (Very Long Baseline Array) at 8 and 15~GHz in 1997 and 2002, respectively. The ridge-line points are indicated by red circles in each image \cite{pe12a}.}
\end{figure}

\section{Radio jets: what do we see?}

    In reference \cite{pe12a} the authors presented a number of observations of the jet in S5~0836+710 obtained at different epochs and different frequencies ranging from 1.6~GHz to 43~GHz, using various ground arrays. The position of the ridge-line along the jet was defined as the peak of emission at a given distance to the radio-core. Figure~1 shows two example images of the jet at 8 and 15~GHz together with the points that form the ridge-lines (red dots). The combination of all the maxima obtained along the peak gives the resulting ridge-line. The jet in S5~0836+710 shows a structure compatible with a projected helix. 
 
    This work showed that the observing array and epoch (through the \emph{uv}-coverage) of a given observation does not have any effect on the large-scale structures observed: The large-scale structures of the ridge-lines of the jet at a given epoch do coincide independently from the observing frequency and array. This result was interpreted as evidence of the physical nature revealed by ridge-lines. Small-scale oscillations may, on the contrary, be produced by observational artifacts. Figure~2 shows the coincidence of the ridge-lines for a given epoch. 
 
   The empirical relationship between the ridge-line and the conditions in the jet that produce more emission was established by the highest-resolution images at 15~GHz (from MOJAVE data, \cite{li09}), which showed that the peak is not necessarily located at the geometrical center of the brightness distribution. On the contrary, the peak of emission is located at the expected positions if it is associated to the pressure asymmetry driving the helical pattern. Therefore, it was concluded that the peak in emission should correspond to a pressure maximum across the jet cross-section.  
      
    In references \cite{co14a,co14b}, Cohen and collaborators presented a similar study for the jet in BL~Lac with observations at 15~GHz and were able to measure the propagation velocity of the ridge-line, which was interpreted as an Alfv\'en wave, under the assumption that the energy flux of the jet is dominated by the Poynting flux.
  
   In addition, the MOJAVE collaboration has shown that the bright features in jets (typically associated to traveling shock-waves) appear moving in different directions from the radio-core with time and that a single observation at a given epoch would not show the complete width of the jet \cite{li13}. The obvious conclusion is that the jet flow fills a wider region than that revealed at a single epoch. A remarkable example of the structure described in this section, albeit at kiloparsec scales, is the \emph{Chandra} image of the jet in the radio-galaxy NGC~315 \cite{wo07}.

    Summarizing the collected information, there is evidence that the emission maxima along radio jets do not correspond to observational artifacts. Furthermore, the jet flows at parsec-scales are wider than the region shown in radio observations. We do not have any conclusive evidence, however, about the exact nature of the observed wave-structures or how the interaction with the propagating perturbation influences them.
         
\begin{figure}[!t]
\center
\includegraphics[width=\textwidth,angle=0,clip=true]{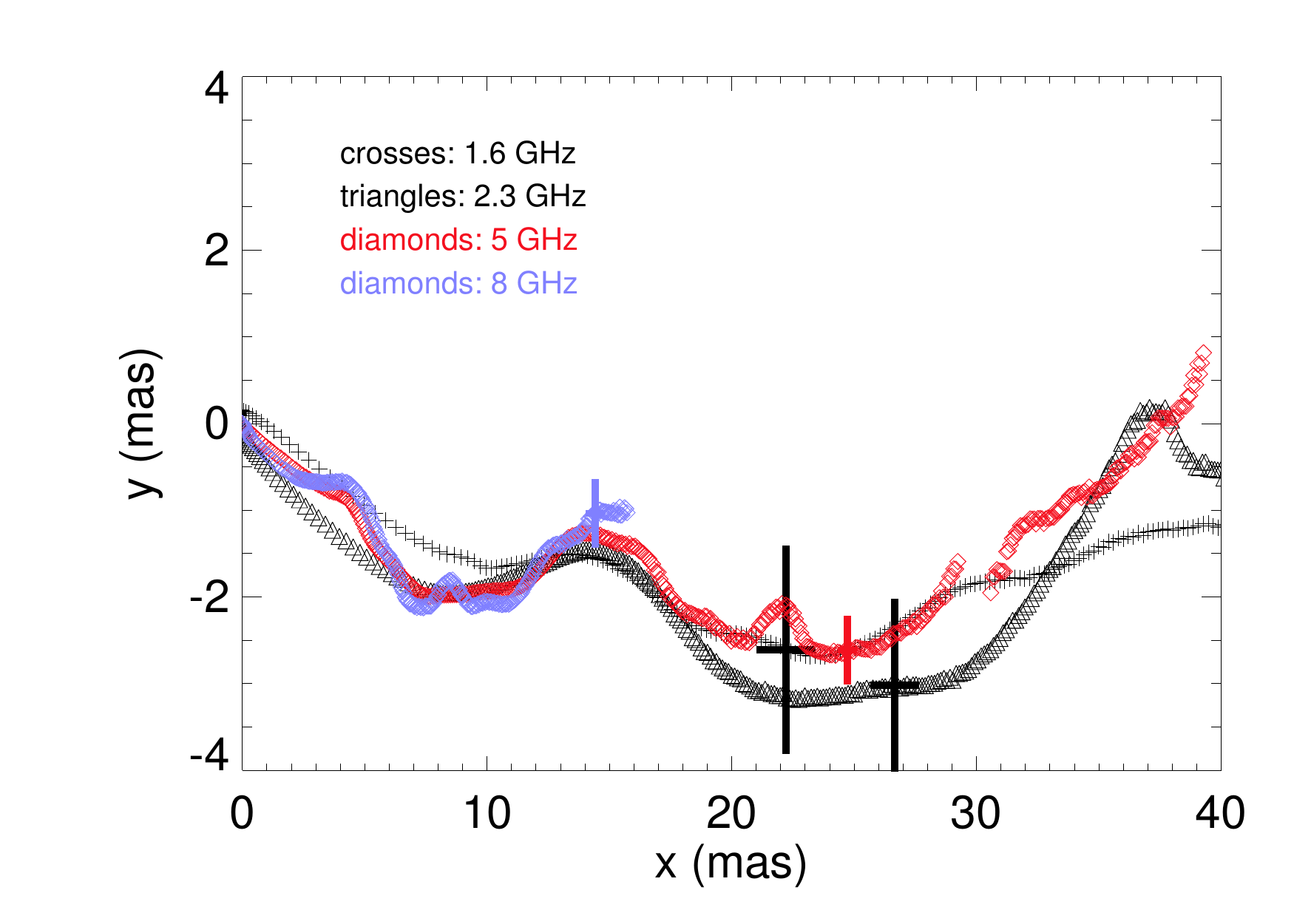} 
\caption{\label{fig1} Ridge-lines over plot of the ridge-lines obtained at different frequencies in 1997 \cite{pe12a}.}
\end{figure}

\section{Discussion}

\subsection{Jet transversal structure} 
   
    The parsec-scale jets seem to show only a portion of their complete cross-section at GHz frequencies. The following discussion from \cite{pe12a} claims that there are two possible explanations for one side of the jet being brighter than the other:

    \emph{If the jet is transversally resolved, the maximum in the emission, or the
position of the ridge-line can be due to two main factors: 1) differential
Doppler boosting caused by velocity gradients in the jet, or 2) an increase in
the local pressure and number of emitting particles. If the position of this maximum is
due to differential Doppler boosting, 
changes in the flow direction would have an influence on the local emission from
the different regions. This should result, depending on the internal velocity
gradient, in irregular structures, contrary to the results obtained. In addition,
relatively large changes in speed and/or flow direction are required in order to
get significant differential Doppler boosting within the jet. Thus, the most
reasonable hypothesis is that the ridge-line corresponds to a pressure maximum.}

   It is possibly important to note that emission asymmetries can also be produced by a helical field. However, this option requires that the magnetic field changes intensity along the jet resulting in a periodical oscillation of the bright region with distance. And this, in turn, brings us again to factor 2 in the previous paragraph: a perturbation in the jet that propagates (rotating) downstream. How large the emissivity difference has to be for our detectors not to reveal the weakly emitting side of the jet remains to be studied.  
  
 % These structures have been explained as bright blobs that follow helical trajectories \cite{bri10}. However, if this was the case, large-scale helical jets as that observed in S5~0836+710 would require that those blobs are all injected with the same direction. Otherwise we could not observe a single projected helix. This statement, of course, does not mean that there cannot be non-ballistic motions of radio-components in jets, which have actually been measured (see, e.g., \cite{li13} and references therein). 

\subsection{Waves or pulses?}

  It has been suggested, for the case of BL~Lac \cite{co14b}, that the observed pattern corresponds to a pulse traveling along the jet magnetic field as an Alfv\'en wave. On the other hand, in the case of 0836+710, it was pointed out that the longest helix should correspond to a periodical perturbation producing a wave-pattern \cite{pe12a}. Jet disruption at larger scales was reported later \cite{pe12b}, and it was interpreted as further evidence of the helical wave coupling to a disruptive KH mode, growing in amplitude and eventually disrupting the flow.   
  These two views are, nevertheless, compatible: Certain jets showing large-scale helical structures can be subject to precession motion at the jet base and, at the same time, irregularities at the formation region can be produced by changes in the accretion rates onto the central compact object (e.g., \cite{ma02}) or any other kind of stand-alone perturbation, e.g., perturbations triggered by asymmetries in the jet itself or in the ambient medium, lateral winds or entrainment of clouds that rotate around the galactic centre from a side of the jet, for instance. These processes are, however, extremely difficult to detect and isolate.

\subsection{Magnetically dominated?}  
  
   As we stated before, in the case of BL~Lac the structures observed to propagate downstream were interpreted as an Alfv\'en wave in a magnetically dominated jet \cite{co14b}. The strongest evidence to favor this hypothesis is the well-ordered distribution of polarization vectors. However, this cannot be a conclusive evidence of a strong magnetic field, taking into account that if the flow of particles is ordered (e.g., collimated by external gas pressure) and dominates the dynamics, similar distributions could be observed under the appropriate circumstances. In addition, different authors have discussed that during the jet acceleration process the magnetic field is the only available energy source for the jet particles to be accelerated (see, e.g, \cite{ko13}). Therefore, stronger evidence is required to clarify this aspect.

\subsection{Interactions between flow perturbations and helical structures.}

   Bright features are routinely observed traveling along parsec-scale jets (e.g., \cite{li13}). These are typically associated to inhomogeneities at injection that trigger shock waves in the jet because they are supersonic. Both these perturbations and helical structures may coexist in jets, as shown by numerical simulations and recent results from the blazar CTA~102 (e.g., \cite{pe06,fr13a,fr13b}). 
   
    On the one hand, jet precession at the jet base can trigger the large-scale morphology in jets as S5~0836+710 \cite{pe12a}, or a perturbation of the magnetic field could propagate downstream in the way that has been reported in BL~Lac \cite{co14b}. On the other hand, perturbations in the mass-energy flux can produce the observed radio-components. These processes can be related to physical processes as different as precession of the rotation axis of the central region and accretion flux, so they could both happen.
   
 %  At parsec-scales, where the traveling radio components have a relatively high brightness, we expect that they completely change the structure of the ridge-lines. A clear signature would be the breaking of the oscillatory pattern that ridge-lines provide in helical jets. This effect is possibly observed at the 15~GHz images of S5~0836+710.
    
\subsection{On transversal motions and the radio-core.}

  The jet in S5~0836+710 showed transversal motion of the ridge-line on the plane of the sky at 15~GHz compatible with an oscillatory behavior, along the first 3~mas of the radio-jet (see Figure~3) \cite{pe12a}. The displacements measured were smaller than the errors in the determination of the positions and the velocities measured were superluminal, which cannot be explained by projection effects (the motions are measured on the plane of the sky). This effect has also been observed in the motion of radio-components in sources like NRAO~150 \cite{ag07,mo14} and OJ~287 \cite{ag12}. 
  
  In particular, for the case of S5~0836+710, it was suggested that the radio-core could be oscillating and that simply by forcing it to fall at the same position introduced artificial displacements when measuring the ridge-line displacements. It was shown that even a very small oscillation (beyond the available resolution) could be responsible for the apparent superluminal motions. Moreover, the oscillation of the radio-core position had been proven for another, closer source (M~81, \cite{mv11}) using the phase-reference technique. The same idea has been recently applied to the superluminal transversal motion of radio-components in NRAO~150 \cite{mo14}.

  \begin{figure}
\center
\includegraphics[width=\textwidth,angle=0,clip=true]{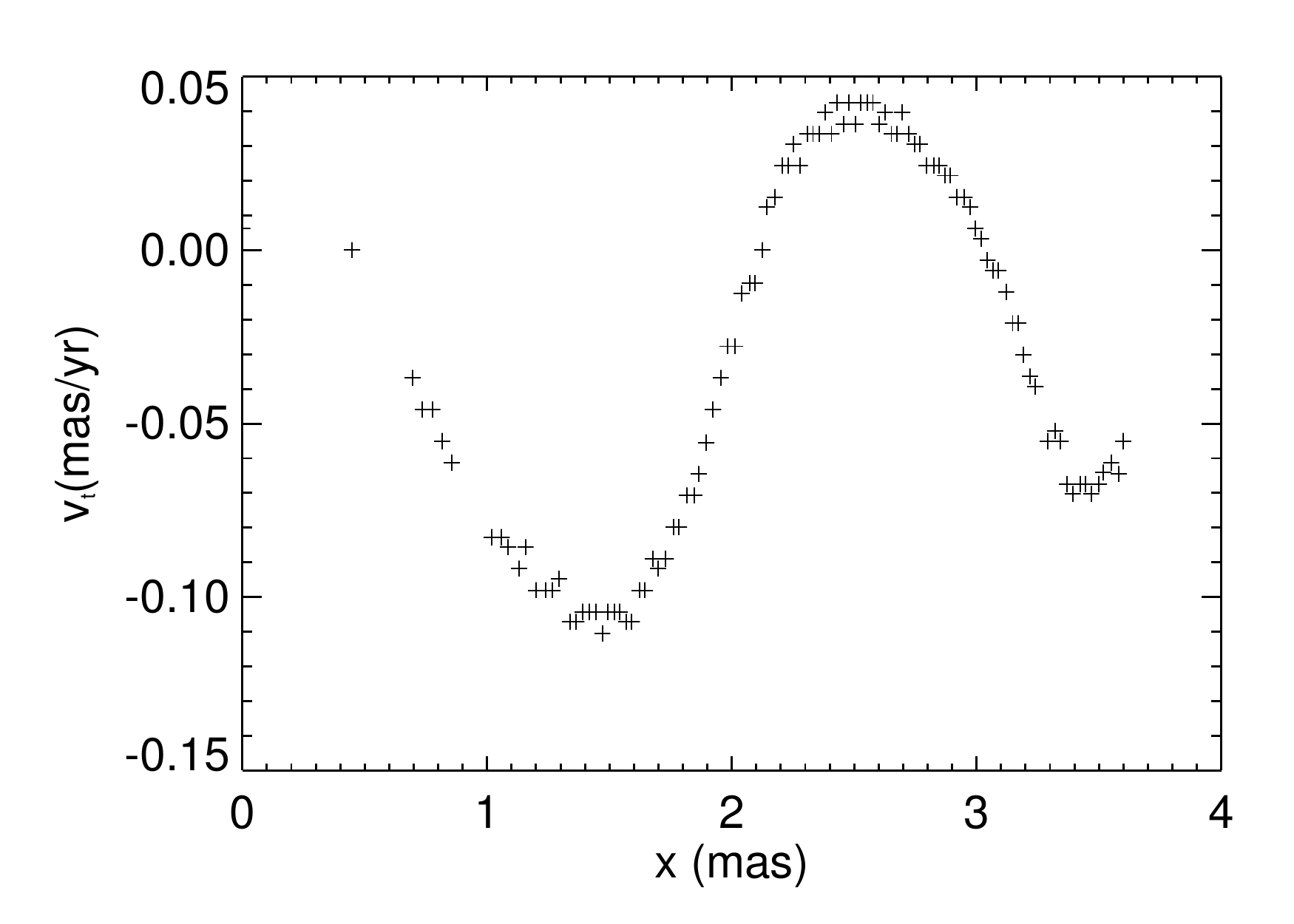} 
\caption{\label{fig3} Measured transversal velocity from two epochs at 15~GHz in mas/yr \cite{pe12a}.}
\end{figure}

\subsection{Prospects from space-VLBI}

  The success of the first observation of the quasar S5~0836+710 using the Russian space antenna \emph{Radioastron} in combination with the available on Earth arrays (Global VLBI) will allow for unprecedented resolution in the determination of the ridge-line location at the observing frequencies: 1.6~GHz and 5~GHz so far. In addition, these observations will be compared with the images obtained in 1997 with VSOP (VLBI Space Observatory Program) using the previously available space-antenna (the Japanese \emph{HALCA}) \cite{lo06}. The aim is to try to measure the transversal velocity of the ridge-line using the displacement between those well resolved images. This measure can help to characterise the structure in detail and obtain relevant information about the jet physics.

\section{Summary}   
  
  Relativistic outflows in active galaxies represent one of the most extreme scenarios that we know in the Universe. They are macroscopic laboratories of plasma physics in the relativistic regime and could play a fundamental role in the evolution of their host galaxy (typically a massive galaxy at the center of clusters) and its environment. Among the open issues in this field we find the relation between the observed radio jets and the bulk (thermal) flow. 
  
   In this contribution, I have summarized the evidence published in recent papers about the association of the brightness ridge-line in radio-jets with wave-like structures. There are, however, different models that could explain them. On the one hand, the helical pattern given by the ridge-line can be interpreted as a wave produced by perturbations at the or close to the formation regions that couple to an unstable KH mode and grow in amplitude as they propagate downstream. Under this interpretation, the ridge-line would correspond to the region with maximum pressure (implying a larger concentration of particles and magnetic field) and would oscillate with time across the jet. On the other hand, the motion of the bright ridge-line can be interpreted as an Alfv\'en wave produced by a perturbation of the magnetic field in a magnetically dominated flow. In addition, the radio components often associated to differential injections of plasma in the jet produce shocks that travel through the jet and may change the brightness configuration of the jet at the affected regions. Altogether, jets are an extremely complex scenario that requires further modeling and observing efforts.  
  
   This complexity requires accurate observations. Thus, only by means of high resolution observations using the space antenna \emph{Radioastron} will we be able to make a step further in the understanding of the physics involved in the processes described in this contribution. The first data obtained from these observations is being currently analysed.

\section*{Acknowledgments}   
I obtained financial support by the Spanish ``Ministerio de Ciencia e Innovaci\'on''
(MICINN) grants AYA2010-21322-C03-01 and AYA2010-21097-C03-01 through an extension of these grants that had expired nine months before the meeting. No new call had been open on time and during 2014 we experienced significant lack of support. I want to thank Y.Y. Kovalev, A.P. Lobanov, P.E. Hardee and I.Agudo for their contribution to this project. I acknowledge Desir\'ee Igual-Chic\'on and Aina Perucho-Igual for their support. 

% Do not delete if you declare acknowledgments
%
%%% ACKNOWLEDGMENTS
%%% ACKNOWLEDGMENTS

%
% Do not delete the next few lines

%
\end{document}